\documentclass[prl,preprint,onecolumn,10pt,letterpaper,superscriptaddress,notitlepage,nobibnotes,longbibliography]{revtex4-1}

\usepackage{times,amsmath,amsfonts,amssymb,latexsym,textcomp}
\usepackage{color}
\usepackage{graphicx}
\usepackage{multirow}
\usepackage{bbm}
\usepackage{tabularx}
\usepackage{amsthm}
\usepackage{dsfont}
\usepackage{placeins}
\usepackage[lofdepth,lotdepth,caption=false]{subfig}
\usepackage[T1]{fontenc}
\usepackage[latin9]{inputenc}
\usepackage[english]{babel}
\usepackage{float}
\usepackage{array}
\usepackage{braket}
\usepackage{MnSymbol}
\usepackage{tikz}
\usepackage[electronic]{ifsym}
\usepackage{booktabs}
\usepackage{bbold}
\usepackage{pdfpages}
\usepackage{pgffor}

\makeatletter
\AtBeginDocument{\let\LS@rot\@undefined}
\makeatother

\usepackage{hyperref}

\renewcommand{\phi}{\varphi}

\renewcommand{\epsilon}{\varepsilon}

\usepackage{fancyhdr}
\begin{document}
\pagestyle{fancy}
\renewcommand\headrulewidth{0pt}
\lhead{}\chead{}\rhead{}
\cfoot{\thepage}

\title{Benchmarks of Nonclassicality for Qubit Arrays}

\author{Mordecai Waegell}
\affiliation{Institute for Quantum Studies, Chapman University, Orange, CA, USA}

\author{Justin Dressel}
\affiliation{Institute for Quantum Studies, Chapman University, Orange, CA, USA}
\affiliation{Schmid College of Science and Technology, Chapman University, Orange, CA, USA}
\email[Corresponding Author: ]{dressel@chapman.edu}
\altaffiliation{Keck Center for Science and Engineering, Chapman University, One University Drive, Orange, CA 92866; Phone: (714) 516-5949}
\maketitle

\pagebreak

\noindent \textbf{Abstract:}  We present a set of practical benchmarks for $N$-qubit arrays that economically test the fidelity of achieving multi-qubit nonclassicality. The benchmarks are measurable correlators similar to 2-qubit Bell correlators, and are derived from a particular set of geometric structures from the $N$-qubit Pauli group. These structures prove the Greenberger-Horne-Zeilinger (GHZ) theorem, while the derived correlators witness genuine $N$-partite entanglement and establish a tight lower bound on the fidelity of particular stabilizer state preparations. The correlators need only $M \leq N+1$ distinct measurement settings, as opposed to the $2^{2N}-1$ settings that would normally be required to tomographically verify their associated stabilizer states. We optimize the measurements of these correlators for a physical array of qubits that can be nearest-neighbor-coupled using a circuit of controlled-$Z$ gates with constant gate depth to form $N$-qubit linear cluster states.  We numerically simulate the provided circuits for a realistic scenario with $N=3,...,9$ qubits, using ranges of $T_1$ energy relaxation times, $T_2$ dephasing times, and controlled-$Z$ gate-fidelities consistent with Google's 9-qubit superconducting chip. The simulations verify the tightness of the fidelity bounds and witness nonclassicality for all nine qubits, while also showing ample room for improvement in chip performance.

\section{Introduction}

As hardware is developed to implement quantum circuits on increasing numbers of qubits, it will be valuable to have economical benchmarks of fully quantum behavior. From the outset of quantum computing it has been clear that the advantage of a quantum computer lies somewhere in its ability to readily perform tasks that are physically challenging or impossible for a classical system. Therefore, ideal hardware benchmarks should certify the ability of the hardware to generate such nonclassical behavior. Indeed, a wide variety of benchmarking techniques have been developed recently \cite{Friis2018,Gheorghiu2019}, including gate-fidelity benchmarks using randomized gate sequences that avoid the state-preparation and measurement errors, and state-preparation benchmarks that certify particular states while avoiding the exponential scaling of state tomography. 

Despite these recent achievements, quantifying the specific nonclassical resources that lead to quantum computational advantage has remained an elusive goal \cite{vedral2010elusive}. Several earlier proposals for suitable measures like entanglement \cite{hill1997entanglement, wootters1998entanglement, wootters2001entanglement, wong2001potential}, Bell-nonlocality \cite{EPR, Bell1, Bell2, greenberger1989going, greenberger1990bell, Bravyi2018}, or quantum discord and its variations \cite{lanyon2008experimental, ferraro2010almost, Modi2012}, proved to be insufficient on their own due to the discovery of algorithmic counter-examples \cite{Bennett1999,Bennett1999b,Meyer2000,dakic2010necessary,Bera2017}. Recent advances suggest a strong connection between quantum advantage and contextuality \cite{KS, galvao2005discrete, howard2014contextuality,wigner1970hidden,klyachko2008simple}, which is a general structural feature of quantum mechanics that subsumes nonlocality. The most pragmatic metric of nonclassical behavior in quantum devices, however, has been the violation of two-qubit Bell inequalities, or similar entanglement witnesses that can apply to few-qubit subsets of a multi-qubit device \cite{Friis2018b}.

In this article we provide a set of practical hardware benchmarks that naturally generalize two-qubit Bell inequality tests to $N$ qubits, based on the Greenberger-Horne-Zeilinger (GHZ) theorem. As with Bell inequalities, our nonclassicality benchmarks use the experimental violation of a classical bound to quantify the nonclassical behavior of the circuit. Beyond quantifying nonclassicality via a bound-violation, these benchmarks also provide tight lower bounds on the fidelities with which particular stabilizer subspaces have been prepared, and thus witness genuine $N$-qubit entanglement for all states that lie within the targeted subspaces. These benchmarks are optimized for testing controllable qubit arrays with nearest-neighbor coupling. As such, we provide efficient circuits for preparing cluster states that maximally violate these benchmarks with controlled-$Z$ entangling gates, using a constant gate depth of 4 (up to hardware-specific decompositions of the controlled-$Z$ gate \cite{Chow2011,Ghosh2013,Martinis2014,Barends2014,kelly2015state,Chow2014}). Though our benchmarks efficiently verify genuine $N$-qubit entanglement using cluster states, many of the benchmarks may be applied to other stabilizer states and we expect similar benchmarks to exist for all stabilizer states. 

The benchmarks we present here generalize earlier work that was experimentally tested with $N=3,4$ photons \cite{greganti2015practical}, where they were compared to previously proposed state-dependent methods for efficiently verifying the fidelity of particular entangled $N$-qubit preparations \cite{Guehne2007,Wunderlich2011}. These prior methods have already been used to verify multi-qubit entanglement in state-of-the-art experiments with 12 qubits \cite{Gong2019} and 18 qubits \cite{Wang2018}, since the exponential scaling required for traditional state tomography is increasingly prohibitive. Notably, for large $N$ our GHZ-based benchmarks produce a tighter preparation-fidelity bound than these existing methods and similarly produce entanglement witnesses with better scaling.   

\section{Results}

\emph{Nonclassicality Benchmarks:---} Our benchmarks consist of measurable correlators that are compared to derived upper bounds; violation of these bounds characterizes nonclassicality. Each such benchmark corresponds to a specific prepare-and-measure circuit on $N$-qubits with $M\leq N+1$ different measurement settings. The $M$ observables form a structure called an ID (also called an \emph{identity product} \cite{W_Primitive}), which is a set of mutually commuting $N$-qubit Pauli operators whose overall product is the $N$-qubit identity, up to a sign. We express an ID as an $M \times N$ table of single-qubit Pauli operators and the identity $\{Z,X,Y,I\}$, labeled $O_{ij}$ with $i=1,...,M$ and $j=1,...,N$. We also define the shortened label $O_i = \bigotimes_{j=1}^{N} O_{ij}$ to indicate the $N$-qubit observable obtained as the product of the $i$th row of an ID. We omit tensor product symbols for compactness.

\begin{figure}[h]
\includegraphics[width=6in]{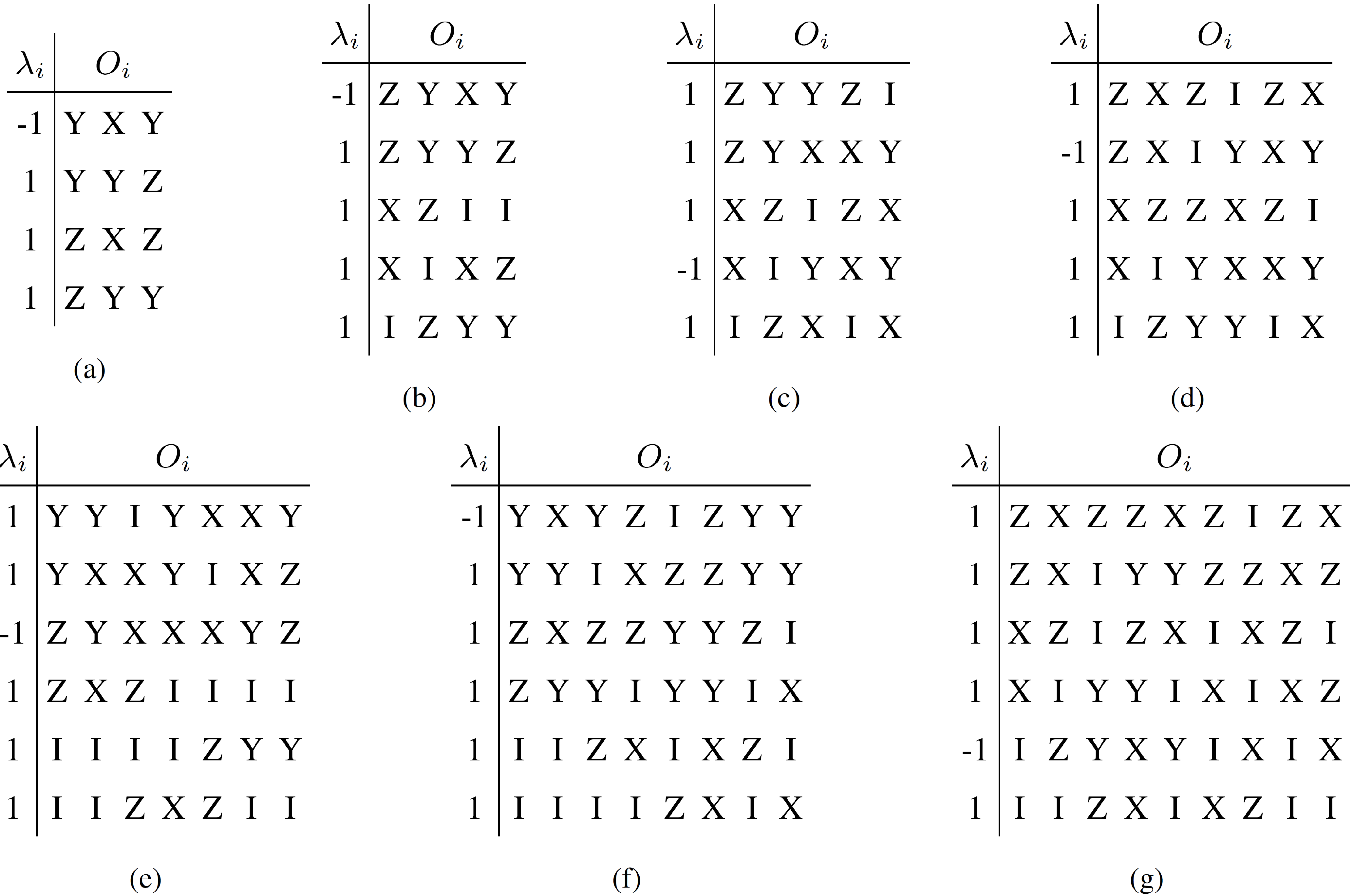}
\caption{Minimal benchmark IDs for $N=3,\ldots,9$ qubits. Each table in (a)--(g) has $M$ rows of $N$ observables $O_{ij}$, with $i=1,\ldots,M$ and $j=1,\ldots,N$. The product of each row defines $O_i = \bigotimes_{j=1}^N O_{ij}$. Eigenvalues $\lambda_i$ of $O_i$ are also shown in each table, chosen to correspond to the state prepared by the circuit of Fig. \protect\ref{Circuits} for the corresponding $N$, which lies in the specific eigenspace stabilized by the ID. Combining the rows of each ID with the appropriate eigenvalue defines a correlator observable $\alpha = \sum_i \lambda_i O_i$, from which we obtain the experimental benchmark score $\mathcal{B} = (\langle\alpha\rangle_\textrm{exp} - M + 2)/2$ that witnesses nonlocal $N$-partite entanglement when $0< \mathcal{B}< 1$, as well as the lower bound $F \geq F_\textrm{ID} = (\mathcal{B} + 1)/2$ on the fidelity $F$ for the state preparation to lie within the indicated eigenspace of the ID.  }
\label{IDs}
\end{figure}

\begin{figure}[h]
\includegraphics[width=5in]{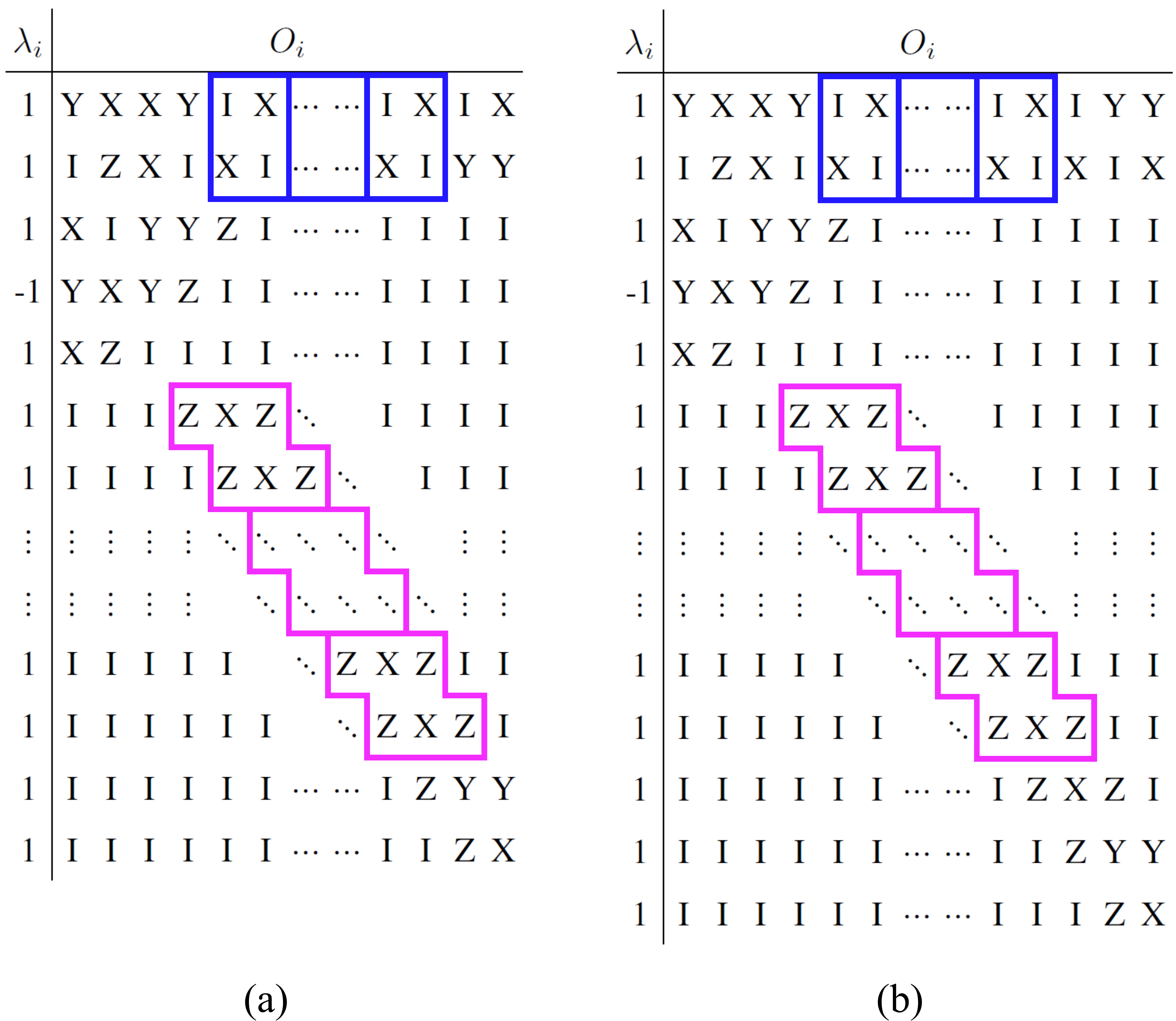}
\caption{Maximal benchmark IDs for (a) all even $N \geq 10$, and (b) all odd $N \geq 11$. These IDs can be extended in increments of two qubits and two observables by adding tiles as shown, and filling all other spaces with `I's.  The $N=10$ and $N=11$ are the cases of (a) and (b), respectively, with zero tiles added.  We can see from the asymmetric shape of the tiles that the added qubits must become entangled with the existing ones because the 2-qubit observables in the added columns do not mutually commute. See the Supplementary Notes and Supplementary Figure 6 for a proof that these IDs belong to the stabilizer group of the linear cluster state for all $N$.}
\label{IDNs}
\end{figure}

To obtain the Bell inequality for each ID \cite{greganti2015practical}, we choose a particular eigenspace $\Pi$ represented by a projector of rank $2^{N-M+1}$, which is specified by the set of $N$-qubit Pauli observables $\{O_i\}$ that form the $M$ rows of the ID (see Figs. \ref{IDs} and \ref{IDNs}), and a specific choice of their respective eigenvalues $\{\lambda_i\}$. We then define the \emph{correlator} observable for this chosen eigenspace,
\begin{equation}
\alpha = \sum_{i=1}^M \lambda_i \, O_i,
\end{equation}
such that its expectation value in a state $\rho$ has an upper bound of $\beta_{\textrm{QM}} = M$, saturated by the chosen eigenspace $\rho = \Pi$,
\begin{equation}
\langle\alpha\rangle = \sum_{i=1}^M \lambda_i\, \textrm{Tr}(\rho\, O_i) \leq \beta_{\textrm{QM}} = M.
\end{equation}
For example, we could prepare the joint eigenstate of the ID of Fig. \ref{IDs}(a), with negative eigenvalue $\lambda_1 = -1$ for the 3-qubit Pauli observable $O_1 = YXY$, and positive eigenvalues $\lambda_2 = \lambda_3 = \lambda_4 = +1$ for the remaining observables $O_2 = YYZ$, $O_3 = ZXZ$, and $O_4 = ZYY$. Then, $\langle\alpha\rangle = \textrm{Tr}(\Pi \alpha) = 4$ since each term in the sum becomes +1.

In the spirit of Bell \cite{Bell1,Bell2}, if one tries to explain the observed correlation by choosing a complete set of local hidden variables $v_{Zj},v_{Xj},v_{Yj} \in \{+1,-1\}$ that predict the outcomes of the single-qubit Pauli measurements, then at least one of the terms in the correlator sum becomes -1, resulting in a smaller upper bound,
\begin{equation}
\langle\alpha\rangle \leq  \beta_{\textrm{LHVT}} = M-2.   \label{Bell}
\end{equation}
Experimental violation of this bound thus indicates nonclassicality in the form of a violation of local realism. Though the locality loophole is always open for neighboring qubits on a chip, this violation is still a useful witness for nonclassical states prepared by the chip, much like for Bell inequalities or Bell-Leggett-Garg inequalities \cite{white2016preserving}. The derivation of this bound is reviewed in the Methods Section.

As an independent result, maximizing the expectation value of the correlator over all biseparable quantum states in the $N$-qubit Hilbert space produces the upper bound,
\begin{equation}
\langle \alpha \rangle \leq \beta_{\textrm{bisep}} = M-2, \label{Witness}
\end{equation}
which happens to coincide with the bound for local hidden variable theories. Experimental violation of the bound thus also witnesses genuine $N$-partite entanglement. In the Methods section, we provide the proof that the joint eigenspaces of the IDs in this article are maximally entangled, as well as the derivation of this bound.

In light of the convenient fact that $\beta_{\textrm{bisep}} = \beta_{\textrm{LHVT}}$, we define the \emph{nonclassicality benchmark score} for a given physical $N$-qubit device as the experimentally determined value,
\begin{equation}
\mathcal{B} = \frac{\langle\alpha\rangle_\textrm{exp} - M + 2}{2},
\end{equation}
such that $\mathcal{B} \leq 0$ fails to witness either entanglement or the violation of local realism, while $0 <\mathcal{B} \leq 1$ witnesses nonlocal $N$-partite-entangled states. The nonclassicality benchmark score thus serves as a metric of uniquely quantum behavior, with $\mathcal{B} = 1$ indicating maximum nonclassicality that saturates the correlator bound. Each $N$-qubit ID provides a benchmark corresponding to a distinct nonclassical eigenspace of an $N$-qubit physical device, and thus the hierarchy of IDs presented in Fig. \ref{IDs} provides a corresponding hierarchy of benchmarks.

\emph{Lower Bounding the Fidelity:---}  The correlator also serves to bound the fidelity from below \cite{greganti2015practical},
\begin{equation}
F \geq F_\textrm{ID} = \frac{\langle \alpha \rangle_\textrm{exp} - M + 4}{4} = \frac{\mathcal{B}+1}{2},   \label{FID}
\end{equation}
where $F = \textrm{Tr}(\rho_\textrm{exp} \Pi) \in [0,1]$ is the fidelity that the experimentally prepared state $\rho_{\textrm{exp}}$ lies within the eigenspace $\Pi$ stabilized by the chosen ID.
We provide a general derivation of this bound in the Methods section.
Importantly, in the limit $\langle \alpha \rangle_\textrm{exp} \to  \beta_{\textrm{QM}} = M$, we have $F_\textrm{ID} \to 1$, and thus as the fidelity of the preparation is improved, this lower bound obviates the need for full tomography of these preparations.

Taken together, the inequalities of Eqs. \ref{Bell}, \ref{Witness}, and \ref{FID} provide a practical and efficient characterization of the prepared $N$-qubit state, as well as a robust benchmark of its nonclassical behavior, using only $M \leq N+1$ measurement settings. 
We present minimal benchmark IDs in Fig.~\ref{IDs} for $N=3,\ldots,9$, and detail minimal IDs up to $N=33$ qubits in Supplementary Figures 1 through 5. These minimal IDs saturate the conjectured bound $N \leq (M-2)(M-1)/2$. We also present a family of maximal benchmark IDs in Fig.~\ref{IDNs} for all $N\geq 10$ that saturate the bound $M -1 \leq N$.

\emph{Benchmark Circuits and Simulation:---}  The IDs in this article have been specially chosen so that the prepare-and-measure circuit for each measurement setting requires a gate depth of 4 on any array of $N$ physical qubits with only nearest-neighbor controlled-$Z$ couplings, making them a scalable and uniform set of benchmarks for implementations of this type. Figure \ref{Circuits} shows the circuits for $N=4,5$, from which the generalization to all $N$ should be straightforward. In general, each circuit prepares an $N$-qubit linear cluster state, which is contained within the maximally entangled subspace of the corresponding ID.

In order to evaluate the usefulness of these benchmarks in real-world physical implementations, we simulated the performance of these circuits for each of the IDs in Fig.~\ref{IDs}. We simulated each circuit over a range of $T_1$ energy relaxation times, $T_2$ dephasing times, and angular jitter for the controlled-$Z$ gate rotations, using the ranges given in Figs.~\ref{Sim1} and \ref{Sim2}. We also considered the effect of initialization and readout error for each qubit. The ranges of values were chosen to match the reported values of the 9-qubit Google chip \cite{Barends2014,kelly2015state}, with the experimental values roughly in the center of each simulated range.  We ran one version of the simulation using a nominal initialization error for each qubit of $P_\textrm{e} = 2\%$, and another version where we used the observed initialization errors for each of the nine qubits on the Google chip. Final readout error has been neglected as correctable for ensemble statistics. Selected plots from the simulations are shown in Fig.~\ref{Sim1}, while scatter plots of the lower fidelity bound, $F_\textrm{ID}$, are shown in Fig.~\ref{Sim2} for the full ranges of simulated values.  Note that in order to minimize the effect of the two worst qubits on the chip (boldface values in Figs. \ref{Sim1} and \ref{Sim2}), we always used the last $N$ qubits on the chip to form our $N$-qubit IDs in the simulation.  See the Methods section for additional details about how the numerical simulations were performed.

Judging by our simulated data shown in Figs.~\ref{Sim1} and \ref{Sim2}, we expect the 9-qubit Google chip to be able to violate the classicality bounds for all nine qubits. We can see clearly that the qubit initialization error is the dominant source of error as we try to move to larger $N$. This shows that our benchmarking scheme is immediately relevant, since it appears that similar hardware fidelity would only violate the bound for one or two more qubits --- but certainly not all 72 on the Bristlecone chip \cite{Note1} --- once suitable IDs have been found beyond the 9 presented here.

\section{Discussion}

The IDs and implementation circuits presented in this article are good benchmark tests for any physical implementation of qubits in a nearest-neighbor-connected array. They work naturally on a chip with more connectivity than this as well. While our simulations targeted a particular recent chip implementation for concreteness, this does not constrain the general usefulness of this protocol for other multi-qubit systems.

\begin{figure}[h]
\includegraphics[width=7in]{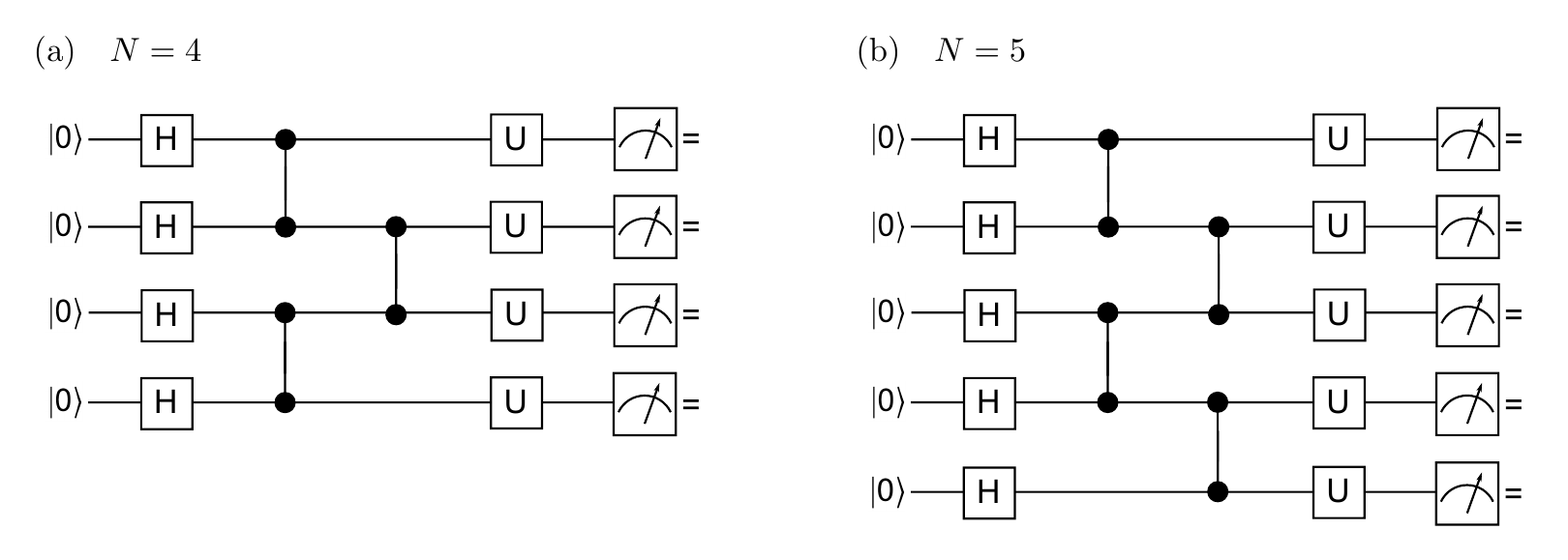}
\caption{Illustrative circuit diagrams for preparing the states for IDs in Fig.~\ref{IDs}, with (a) corresponding to $N=4$ in Fig.~\ref{IDs}b and (b) corresponding to $N=5$ in Fig.~\ref{IDs}c. These two examples generalize to $N$-qubits, and produce linear cluster states. The local measurement settings for each observable $O_{ij}$ in the ID are implemented by the unitary operations $U_{ij}$, assuming detectors that naturally measure the $Z$ basis. This circuit allows the $M$ different settings of an ID to be implemented with different $U_{ij}$ for different observables and qubits. For example, in the 5-qubit ID of Fig. \protect\ref{IDs}c the first setting is $ZYYZI$, meaning that for the first and fourth qubits $U_{11}= U_{14}=I$, for the second and third qubits $U_{12} = U_{13} =e^{i\pi X /4}$, and the fifth qubit is ignored.}
\label{Circuits}
\end{figure}

\begin{figure}[h]
\includegraphics[width=7in]{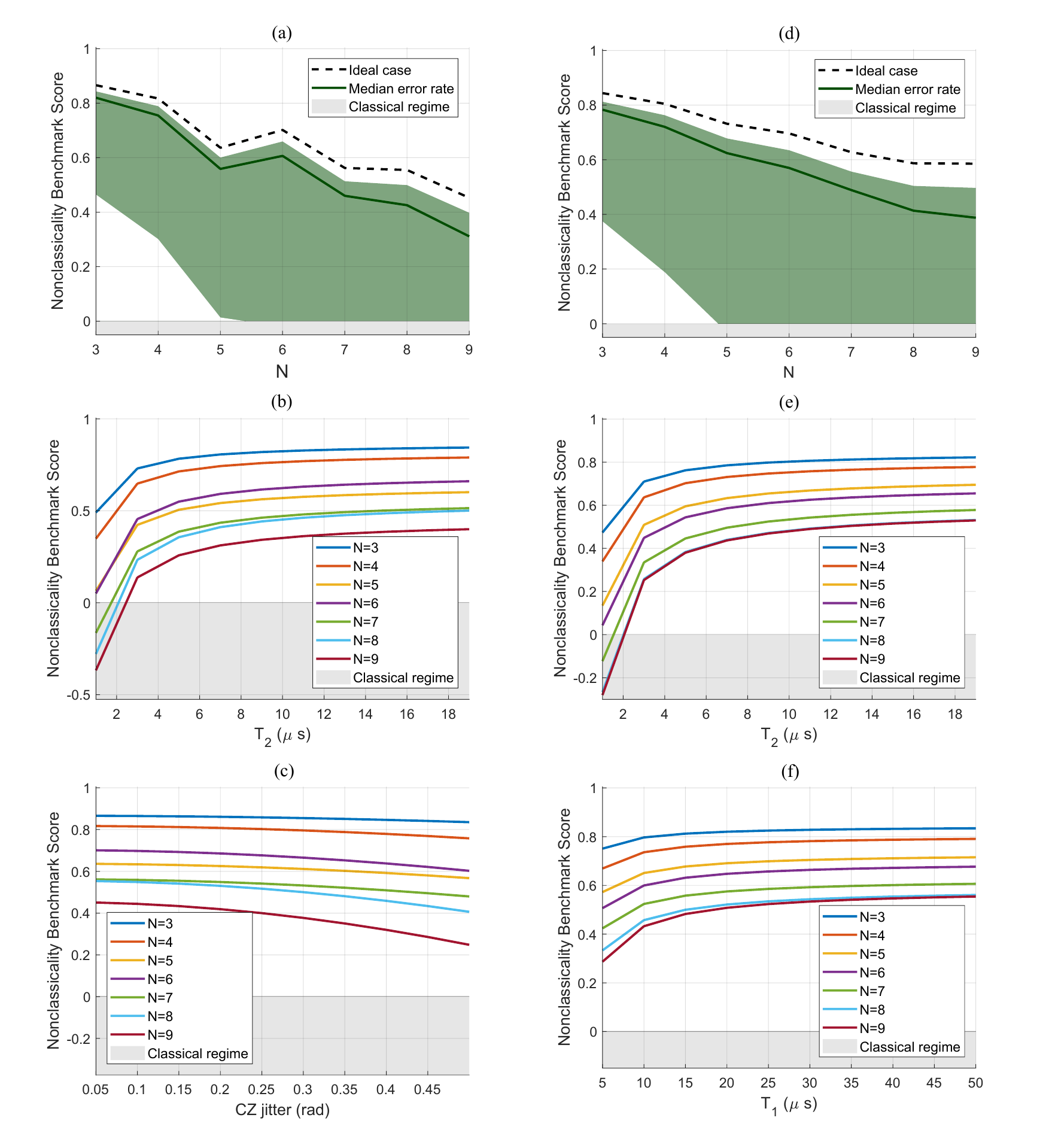}
\caption{Nonclassicality Benchmark Scores ($\mathcal{B}$), for selected simulations. The nonideality parameter ranges were $T_1\in[5,50]$ $\mu$s energy relaxation times, $T_2\in[1,19]$ $\mu$s dephasing times, and $w\in[0.05,0.5]$ rad of angular jitter widths for $ZZ_{90}$ decompositions of controlled-$Z$ (CZ) gates. We used a single-qubit gate time of $\Delta t = 25$ ns, and a 2-qubit controlled-$Z$ gate time of $\Delta t = 45$ ns, which are conservative estimates for the reported gate times. In plots (e) and (f), the curves for qubit numbers $N=3,\ldots,9$ are ordered starting from the top curve. In plots (b) and (c), the $N=5$ lines are below the $N=6$ lines, and the $N=7$ lines are below the $N=8$ lines, due to the poor performance of chip qubits 5 and 7 (boldface values).  (a,b,c) Simulated data using Google's 9-qubit-chip values \cite{Barends2014,kelly2015state}: $\{T_1\} = \{   18.6,   28.1,  22.0,$ $    19.1,   \textbf{41.1},   21.3 ,  \textbf{39.2} ,  24.7 ,  26.3\} \mu $s and $\{P_e\} = \{ 1.8,    1.1 ,   1.7,    1.3  ,  \textbf{4.8}   ,  0.7 ,   \textbf{6.7}  ,   0.4  ,  1.5 \} \% $. (d,e,f) Simulated data for $P_e = 2\%$ initialization error, with parameter ranges centered on mean chip values. (a,d) $\mathcal{B}$ vs. $N$. Ideal curves have $T_2 = T_1 = \infty$ and $w=0$. Median curves approximate the chip, with shading indicating the range of simulated values.  (b,e) $\mathcal{B}$ vs. $T_2$, fixing median chip values of $w$ and $T_1$. (c) $\mathcal{B}$ vs. $w$, fixing the median chip value of $T_2$. (f) $\mathcal{B}$ vs. $T_1$, fixing the median chip values of $T_2$ and $w$.}
\label{Sim1}
\end{figure}

\begin{figure}[h]
\includegraphics[width=7in]{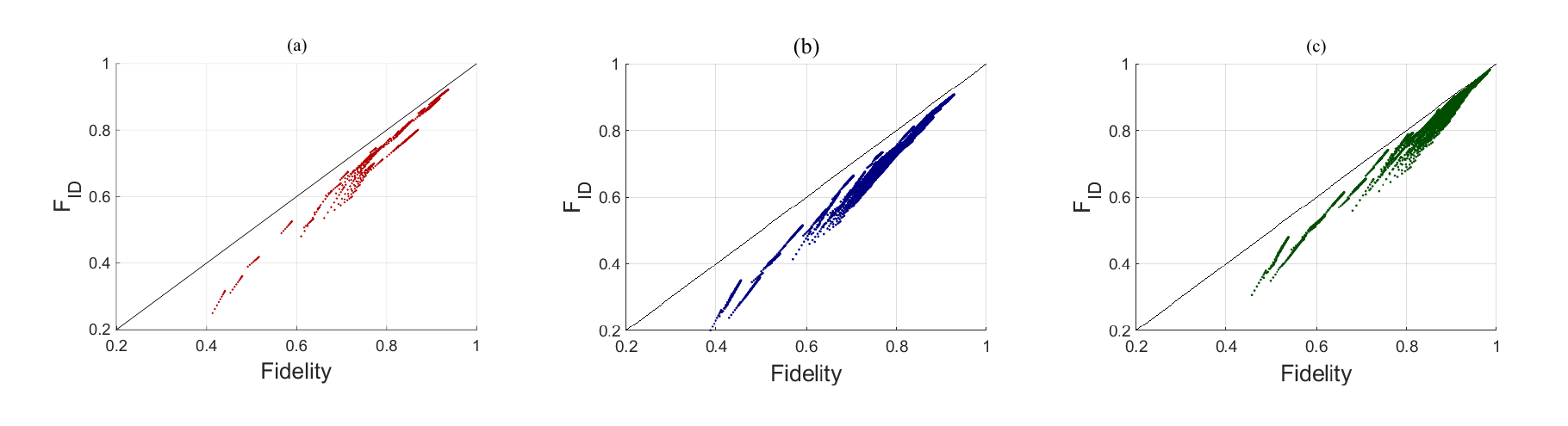}
\caption{Scatterplots of the fidelity lower bound $F_\textrm{ID}$ vs. true fidelity $F$ for all simulated data. The lower bound is tight, thus as $F\to 1$ so too does $F_\textrm{ID}$. All plots contain data for the nonideality ranges: $T_2\in[1,19]$ $\mu$s dephasing times, and $w\in[0.05,0.5]$ rad angular jitter widths for CZ gates. (a) Chip values $\{T_1\} = \{  18.6,  28.1, 22.0,$ $  19.1,  \textbf{41.1},  21.3 ,  \textbf{39.2} ,  24.7 ,  26.3\} \mu $s energy relaxation times, and $\{P_e\} = \{ 1.8,    1.1 ,   1.7,    1.3  ,  \textbf{4.8}   ,  0.7 ,   \textbf{6.7}  ,   0.4  ,  1.5 \} \% $ initialization error. (b) $P_e = 2\%$ initialization error, with range $T_1\in[5,50]$ $\mu$s. (c) Same ranges as the center plot, but with $P_e = 0$ to show the asymptotic approach $F_{ID} \to F$ as $F\to 1$. }
\label{Sim2}
\end{figure}

Although some other families of IDs with the same properties as those in Figs. \ref{IDs} and \ref{IDNs} are known \cite{W_Primitive, waegell2013nonclassical}, the minimal IDs, with the largest possible value of $N$ for a given $M$, are not known in general (see the Supplementary Discussion and Supplementary Figures 1 through 5 for the best known cases). Because of their geometric nature, enumerating all of the representative IDs for given values of $N$ and $M$ is a highly nontrivial problem, related to solving the graph isomorphism problem on $N \times M$ colored vertices, and it is thus limited by computational resources. Furthermore, not every ID can be constructed using only nearest-neighbor couplings in linear circuits as in Fig. \ref{Circuits}. The increased connectivity of more modern chips, like the Bristlecone chip from Google, should allow the use of more general IDs, although the circuit depth will likely increase by one or two gates. 

Each of the IDs presented here also gives rise to a complete proof of the Kochen-Specker (KS) theorem for contextuality \cite{KS,waegell2012proofs,waegell2013proofs}, which can be implemented for any initial state with a few alternative circuits for the different measurement contexts. In general, IDs are the natural building blocks of proofs of the KS theorem in the $N$-qubit Pauli group. This is a slightly more complicated setup, which could inspire different contextuality-based benchmarks in future work.

Finally, maximally entangled IDs with $M < N+1$ give rise to maximally entangled eigenspaces, each of dimension $2^{N-M+1}$, which generalize the codespaces of error correcting codes \cite{knill1997theory, nielsen2010quantum}, and $L = N-M+1$ is the number of \emph{logical qubits} (where $N$ is the number of \emph{physical qubits}). All $N$-qubit-stabilizer-based error correcting codes (including the toric code \cite{kitaev2003fault}) belong to the family of IDs, and while all IDs of this type are error-detecting codes, they cannot all be used to diagnose the syndrome of an error in order to correct it.  Many of the well-known error correcting codes generate an ID which proves the GHZ theorem, and all can be used as entanglement witnesses in the manner of this article \cite{divincenzo1997quantum}. Nevertheless, these more general maximally entangled subspaces may be of significant interest for other applications in quantum information processing, which warrants further investigation. One straightforward application for these subspaces is to perform benchmarks that measure the physical qubits as described in this paper, while simultaneously benchmarking the performance of the logical qubits in some additional way. The two tests may be performed simultaneously because any general logical $L$-qubit state can be prepared for each benchmark, although the circuit is likely to be longer and more complex than Fig. \ref{Circuits}, and the performance will be commensurately worse.  It is remarkable to note that if the conjectured bound $N \leq (M-2)(M-1)/2$ can be saturated, then the number of logical qubits is bounded by $L \leq ( (M-2)(M-1)/2 - M + 1$, and thus the ratio $L/N \rightarrow 1$ in the limit $M \rightarrow \infty$.

\section{Methods}

\emph{Proving the GHZ Theorem:}  All of the IDs in Fig. \ref{IDs} have sign -1, and for each qubit $j$, the number of entries $O_{ij}=Z$ in the ID is even, as is the number of entries with $O_{ij}=X$ and with $O_{ij}=Y$.  These properties indicate that these IDs give rise to proofs of the GHZ theorem \cite{greenberger1989going}, which is a logical version of Bell's nonlocality theorem \cite{Bell1, Bell2}, without any inequalities.  To see this, suppose that a joint eigenstate (i.e., any state in a joint eigenspace) of these observables is prepared.  This eigenstate has $M$ eigenvalues $\lambda_i$ corresponding to the $M$ observables, and $\prod_{i=1}^M \lambda_i = -1$, since the product of these $M$ observables is $-I^{\otimes N}$.  Suppose that each of the $N$ qubits are now mutually space-like separated, and each is subjected to random local Pauli measurements, and label their outcomes $\lambda_{ij}$, when all $N$ local measurement settings happen to correspond to observable $i$ of the ID.  The entanglement correlations that are obeyed by this state are $\prod_{j=1}^N \lambda_{ij} = \lambda_i$.  Putting these relations together we have $\prod_{i=1}^M\prod_{j=1}^N \lambda_{ij} = -1$.  Now, in order for a local hidden variable theory (LHVT) to explain these entanglement correlations, each qubit $j$ must carry local hidden variables $v_{Zj},v_{Xj},v_{Yj} \in \{+1,-1\}$ which predict the outcomes $\lambda_{ij}$, and are pre-arranged to satisfy the entanglement constraints.  However, for such hidden variables we would have $\prod_{i=1}^M\prod_{j=1}^N \lambda_{ij} = \prod_{j=1}^N v_{Zj}^{n_j} v_{Xj}^{m_j} v_{Yj}^{l_j} = +1$, since $n_j$, $m_j$, and $l_j$ are all even for the IDs of this article, and thus is is impossible to choose local hidden variables which can satisfy the entanglement correlations of this state.  This logical proof without inequalities can be converted into a Bell inequality for use as a benchmark of $N$-qubit nonlocality, as shown in the main text, by noting that for any complete assignment of local hidden variables $v_{Zj},v_{Xj},v_{Yj} \in \{+1,-1\}$ to the ID, at least one of the observables has the wrong eigenvalue.

In general, proving the GHZ theorem does not prove that nonlocal correlations exist between more than just a single pair of qubits among the $N$ \cite{collins2002bell,svetlichny1987distinguishing, seevinck2002bell, mitchell2004conditions}, nor does it generally witness genuine $N$-qubit entanglement. In contrast, the benchmark IDs we present in this article prove the GHZ theorem \emph{and} are constructed to be $N$-partite entanglement witnesses \cite{toth2005detecting, toth2005entanglement}, such that their corresponding Bell inequalities can only be violated by genuinely $N$-qubit-entangled states. To go further than the results we present here and prove nonlocal correlations exist between every pair of qubits among the $N$, one must violate the corresponding Svetlichny inequalities \cite{collins2002bell,lavoie2009experimental} instead, but with the cost that the number of required measurement settings grows exponentially with $N$ \cite{collins2002bell}.

\emph{Bounding the Fidelity:}
An $N$-qubit ID with $M$ observables $\{O_i\}$ has a complete set of eigenspaces $\{\Pi_k\}$ satisfying $\sum_k \Pi_k = I$, each of which can be identified by a unique set of distinct eigenvalues $\{\lambda_{ik}\}$ of $\{O_i\}$.  Only $M-1$ of the observables in an ID are independent, and if $M-1<N$ the eigenspaces $\Pi_k$ are degenerate, and each contains $2^{N-M+1}$ mutually orthogonal vectors $|\kappa_{jk}\rangle$ which share the eigenvalue $\lambda_{ik}$, with $j = 1,\ldots,2^{N-M+1}$, such that $\{|\kappa_{jk}\rangle\}$ is a complete orthonormal eigenbasis of the ID.   Each of the $2^{M-1}$ eigenspaces $\Pi_k$ corresponds to a unique correlator $\alpha_k = \sum_{i=1}^M \lambda_{ik} O_i$. Each experimentally obtained quantity $\langle \alpha_k \rangle$ enables us to put a lower bound on the fidelity that an experimentally prepared pure state $|\psi\rangle$ lies within the eigenspace $\Pi_k$ \cite{greganti2015practical}.

With no loss of generality, we will henceforth use correlator $\alpha_1$ and the target eigenspace $\Pi_1$.  We begin by expanding $|\psi\rangle$ in this eigenbasis as,
\begin{equation}
|\psi\rangle =  \sum_{j=1}^{2^{N-M+1}}\Bigg[  a_j|\kappa_{j1}\rangle + \sum_{k=2}^{M-1} b_{jk} |\kappa_{jk}\rangle\Bigg], 
\label{fid0}
\end{equation}
such that $\sum_j\big(|a_j|^2 + \sum_{k=2}^{2^{M-1}} |b_{jk}|^2\big) = 1$.

Since the expansion is in an eigenbasis of $\alpha_1$, we find
\begin{equation}
\langle \alpha_1 \rangle_\textrm{exp} = \langle \psi | \alpha_1 | \psi \rangle = \sum_{j=1}^{2^{N-M+1}}\Bigg[  |a_j|^2\langle\kappa_{j1}|\alpha_1|\kappa_{j1}\rangle + \sum_{k=2}^{2^{M-1}} |b_{jk}|^2 \langle\kappa_{jk}|\alpha_1|\kappa_{jk}\rangle\Bigg].
\label{fid1}
\end{equation}

Note that $\langle \kappa_{j1} | \alpha_1 | \kappa_{j1} \rangle = \sum_{k=1}^M \lambda_k^2 = M$, since all eigenvalues of $|\kappa_1 \rangle$ match those in the correlator $\alpha_1$ by construction, and thus square to 1. However, any other $|\kappa_{jk} \rangle$ does not lie within $\Pi_1$, so is characterized by eigenvalues distinct from those characterizing $\Pi_1$. Moreover, since the product of all eigenvalues for the observables of a given ID is fixed for any eigenstate, only even numbers of eigenvalues can differ from those characterizing $\Pi_1$, which necessarily causes at least two terms of $\langle \kappa_{jk} | \alpha_1 |\kappa_{jk} \rangle$ to become $-1$, resulting in an upper bound of $\langle \kappa_{jk} | \alpha_1 | \kappa_{jk} \rangle \leq M-4$ for those eigenstates. Using these two observations we obtain,
\begin{equation}
\langle \alpha_1 \rangle_{\rm exp} \leq \sum_{j=1}^{2^{N-M+1}}\Bigg[|a_j|^2M + \sum_{k=2}^{2^{M-1}} |b_{jk}|^2 (M-4)\Bigg] = 4F + M-4 ,
\end{equation}
where $F = \sum_j |a_j|^2$, and we have used $\sum_j\big(|a_j|^2 + \sum_{k=2}^{2^{M-1}} |b_{jk}|^2\big) = 1$. We can rewrite this relation as
\begin{equation}
F \equiv \langle \psi | \Pi_1 | \psi\rangle \geq \frac{\langle \alpha_1 \rangle_{\rm exp} - M + 4}{4} \equiv F_\textrm{ID}.
\label{FidBound}
\end{equation}
Noting that the left hand side of this equation is the fidelity $F$ for the preparation $|\psi\rangle$ to lie within the eigenspace $\Pi_1$, the right hand side $F_\textrm{ID}$ gives a lower bound $F \geq F_\textrm{ID}$ for the fidelity. For IDs with $M=N+1$, the target subspace $\Pi_1$ contains only one eigenvector, so the fidelity $F$ is also a state preparation fidelity for the particular target eigenstate $|\kappa_1\rangle$. For IDs with $M<N+1$, the target subspace $\Pi_1$ is degenerate, so the fidelity $F$ is the fidelity for $|\psi\rangle$ to lie within that subspace.

Next we generalize the above derivation to the case of mixed states.  For a general convex combination of $m$ pure states,
\begin{equation}
\rho = \sum_{j=1}^m c_l |\psi_l\rangle\langle\psi_l|,
\end{equation}
where $\sum c_l = 1$, we can expand each $|\psi_l\rangle$ using appropriate eigenbases of the ID as in Eq.~(\ref{fid0}) and follow the same arguments to obtain
\begin{equation}
\langle \alpha_1 \rangle_\textrm{exp} \leq \sum_{l=1}^m c_l (4F_l+M-4),
\end{equation}
where $F_l \equiv \langle \psi_l | \Pi_1 | \psi_l\rangle$.  We can rewrite this as,
\begin{equation}
F \equiv \text{Tr}(\rho\,\Pi_1) = \sum_{j=1}^m c_l F_l \geq \frac{\langle \alpha_1 \rangle_\textrm{exp} - M+4}{4} \equiv F_\textrm{ID}.
\end{equation}
As in the pure state case, the left hand side is the fidelity $F$ for the mixed state $\rho$ to lie within the target subspace $\Pi_1$, while the same expression for the right hand side $F_\textrm{ID}$ places a lower bound on this fidelity.

\emph{Witnessing Genuine $N$-Partite Entanglement:} An $N$-qubit ID provides an entanglement witness if it is maximally entangled \cite{W_Primitive, waegell2014bonding}. Entanglement is usually discussed in reference to the separability of states. However, there is a way to reason about the entanglement of a set of observables directly without reference to states. We define a maximally entangled set of $N$-qubit observables as one with the property that there exists no bipartition of the $N$ qubits into subsets of $R$ and $N-R$, such that all of the observables in each subset $\bigotimes_{k \in [1,R]} O_{ik}$ mutually commute.  It follows from this definition that the joint eigenstates of this set are maximally entangled $N$-qubit stabilizer states.

To see this, consider that every stabilizer state (space) of $N$ qubits has a stabilizer group of $b = 2^g$ mutually commuting Pauli observables $\{S_i\}$ and corresponding eigenvalues $\{\lambda_i\}$, and its density operator can be written as,
\begin{equation}
\rho = \frac{1}{d} \sum_{i=1}^{b} \lambda_i S_i,
\end{equation}
where $g$ is the number of independent generators in the set, and $d=2^N$ is the dimension of the Hilbert space.  Note that if $g<N$, then $\rho$ projects onto a subspace of rank $r = 2^{N-g} > 1$, and that $g = M-1$ for a minimal ID, which is just a specific subset of one or more complete stabilizer groups.  If a stabilizer state is the tensor product of two smaller stabilizer states on subsystems $A$ and $B$, it follows that its density operator can be written as,
\begin{equation}
\rho_{AB} = \Big(\frac{1}{d^A} \sum_{i=1}^{b^A} \lambda^A_i S^A_i\Big) \otimes \Big(\frac{1}{d^B} \sum_{j=1}^{b^B} \lambda_j^B S^B_j\Big) =  \frac{1}{d^{AB}}\sum_{k=1}^{b^{AB}} \lambda^{AB}_k S^{AB}_k.
\end{equation}
For the bipartition of the system into $A$ and $B$, all of the stabilizer operators $S^A_i = \bigotimes_{k \in A} O_{ik}$ mutually commute by definition.  It follows that one can find such a mutually commuting bipartition for any separable state, and therefore if no such bipartition exists, then the set of observables is maximally entangled.  All of the IDs presented in this article are maximally entangled in this way, which results in a witness inequality with the same bound as the Bell inequality.

All states within a maximally entangled eigenspace of an ID are maximally entangled, meaning that for all of them, the maximum squared-Schmid-coefficient across all bipartitions is $1/2$.  For such an eigenstate $|\psi\rangle$, a standard entanglement witness is $\mathcal{W} = \mathbb{1}/2 - |\psi\rangle\langle\psi|$, and an experimental measurement of $\langle \mathcal{W} \rangle < 0$ is a witness of genuine $N$-partite entanglement \cite{toth2005entanglement}.  Noting that a superposition state $a|\psi\rangle + b|\psi_\bot\rangle$ can only violate this bound for $F = |a|^2 > 1/2$, we obtain $F_\textrm{ID} \leq F \leq 1/2$ for all biseparable states.  Plugging this into $F_\textrm{ID} = (\langle \alpha \rangle_\textrm{exp} - M + 4)/4$ yields $\langle \alpha \rangle_\textrm{bisep} \leq M-2$, which is Eq.~(\ref{Witness}).

\emph{Numerical Simulation Details:}
In the simulation, the state is first degraded by initialization error. That is, ideally the $N$ qubits are prepared in an initial ground state $\otimes_{i=1}^N|0\rangle$. However, each qubit has an error probability $P_{\rm e}^{(i)}$ of being initially excited, which produces a mixed initial bit state $(1-P_{\rm e}^{(i)})|0\rangle\langle 0 | + P_{\rm e}^{(i)}|1\rangle\langle 1| = (1-2P_{\rm e}^{(i)})|0\rangle\langle 0| + P_{\rm e}^{(i)} I$, and thus a degraded initial state $\rho = \otimes_{i=1}^n[ (1-2P_{\rm e}^{(i)})|0\rangle\langle 0| + P_{\rm e}^{(i)} I ]$ with ground state fidelity $\prod_{i=1}^n(1-2P_{\rm e}^{(i)})$. The final readout error for an ensemble average can be corrected if the readout misidentification probabilities $P_{\rm e}^i$ are known, and thus we have neglected the role of the readout error.

Each gate in Fig. \ref{Circuits} is then applied to the initial state $\rho$. For the Hadamard gate, it is sufficient to use a $Y_{90}$ rotation, $\exp(-iY\pi/4)$. We decompose the controlled-$Z$ gate into an implementable $ZZ_{90}$ entangling gate and single-qubit corrections: $\exp(i\pi/4)[\exp(iZ\pi/4)\otimes\exp(iZ\pi/4)]\exp(-iZZ\pi/4)$. We degraded each gate by $T_1$ energy relaxation and $T_2$ dephasing processes for the corresponding gate times $\Delta t$. For the energy relaxation time $T_1$, the first-order corrections for each individual qubit are accumulated and then applied to $\rho$. For each qubit $\Delta \rho_i = \big(a^\dagger_i \rho a_i - \frac{1}{2}\{\rho,a^\dagger_i a_i\} \big) \Delta t / T_1^i$, where $a_i$ is the lowering operator of the $i$th qubit tensored with identity for the other qubits, and $\rho \rightarrow \rho + \sum_i^N \Delta \rho_i$. This linear-order Lindblad-form update is sufficient since $\Delta t/T_1^i \ll 1$. For the dephasing time $T_2$, we directly construct the matrix,
\begin{equation}
D = \left(
  \begin{array}{cc}
    1 & e^{-\Delta t / T_2} \\
    e^{-\Delta t / T_2} & 1 \\
  \end{array}
\right)^{\otimes N},
\end{equation}
for efficiency and apply gate dephasing using element-wise multiplication (MATLAB syntax .*), as $\rho \to \rho$ .* $D$.

For simulating gate infidelity, we assume that the single-qubit gate fidelities are high enough for their errors to be neglected, and so simulate only a range of fidelities for the 2-qubit controlled-$Z$ gates. As a crude model for infidelity of a controlled-$Z$ gate, we add a random angular jitter $\delta \phi$ only to the $ZZ$ rotation step, $\exp[-i ZZ (\pi/2+\delta \phi)/2]$, and average over the effect of this jitter using a raised cosine distribution with a width $w$, $dP(\delta \phi) = d(\delta\phi)[1 + \cos(\pi \delta \phi/w)]/(2w)$, where $\delta\phi \in [-w, w]$ has compact angular support. This yields the averaged state update,
\begin{equation}
\rho \rightarrow \int e^{-i\zeta_i(\pi/2 + \delta \phi)/2}\,\rho\, e^{i\zeta_i(\pi/2 + \delta\phi)/2}\,dP(\delta \phi) = \frac{1}{2}\bigg[\rho + \zeta_i \rho \zeta_i  +  i(\zeta_i\rho - \rho \zeta_i)\bigg(\frac{\sin w }{w} - \frac{\sin w}{2(w+\pi)} - \frac{\sin w}{2(w-\pi)}\bigg)\bigg],
\end{equation}
where $\zeta_i$ is the tensor product of Pauli $Z$ for the two qubits the controlled-$Z$ is acting on, and identity for all of the other qubits. The limit as $w\to 0$ restores the unperturbed gate. This crude error model includes only one possible physical mechanism of infidelity for the controlled-$Z$ gate, but gives an indication of the gate sensitivity to imprecise angular control. Since the initialization error dominates the infidelity, the effect of the angular jitter is small.

\textbf{Code availability:} The MATLAB code used to generate our data is available from the authors on reasonable request.

\textbf{Data availability:} The data that support our findings are available from the authors on reasonable request. 

\textbf{Acknowledgments:} We thank Josh Mutus and Daniel Sank for helpful commentary, as well as Eric Freda for helping to create some of the figures in this paper.  MW was partially supported by the Fetzer Franklin Fund of the John E. Fetzer Memorial Trust. JD was partially supported by the Army Research Office (ARO) grant No. W911NF-15-1-0496, as well as No. W911NF-18-1-0178.

\textbf{Competing Interests:} There are no competing interests.

\textbf{Author Contributions:}  MW developed the IDs for the benchmarks in this article, developed the benchmark inequalities, coded the simulations, and wrote the manuscript. JD developed the theory for the simulations, and co-wrote the manuscript.

\section{Figures}

\clearpage



\section*{Supplementary material}

# Benchmarks of Nonclassicality for Qubit Arrays: Supplementary Information

Mordecai Waegell[1] and Justin Dressel[1,2]

[1]*Institute for Quantum Studies, Chapman University, Orange, CA, USA*
[2]*Schmid College of Science and Technology, Chapman University, Orange, CA, USA*[*]

## SUPPLEMENTARY DISCUSSION

**Minimal Benchmark IDs:** We have discovered benchmark IDs which appear to be minimal for all $3 \le N \le 33$, and an example for each value of $N$ is shown in Supplementary Figures 1, 2, 3, 4, and 5. At the time of writing, these IDs have the fewest measurement settings for $N \le 33$, while the maximal class in Fig. 2 of the main text provides a benchmark for all $N \ge 10$. Numerical searches for minimal IDs with higher $N$ are underway, and the largest value of $N$ we can obtain for a minimal ID will be limited by computational resources and time. The largest resource cost comes from checking each ID for maximal entanglement, which generally requires examination of every bipartition by brute force.

The maximal $N$ for known benchmark IDs of a given $M$ are,

| $N$ | 3 | 6 | 10 | 15 | 21 | 28 |
|---|---|---|---|---|---|---|
| $M$ | 4 | 5 | 6 | 7 | 8 | 9 |

This suggests that the number of qubits $N$ in a benchmark ID with $M$ observables is bounded by,

$$M - 1 \le N \le (M-2)(M-1)/2, \tag{1}$$

and we conjecture that this will hold for all $M$. If we can derive a proof that this conjecture is true, this may help to identify a pattern to generate IDs with $N = (M-2)(M-1)/2$ for all $M$.

## SUPPLEMENTARY NOTES

**Maximal Benchmark IDs for All $N$:** The benchmark IDs for all $N$ in the main text belong to the stabilizer group of the $N$-qubit linear cluster state. That is, these maximal IDs stabilize a particular maximally entangled eigenstate, which is rank 1 as opposed to the code spaces of maximum rank stabilized by the minimal IDs. Unlike the case with minimal IDs, we exploit an emergent pattern in the maximal IDs that allow them to be extended to all $N$ in a straightforward way that is provably maximally entangled.

To prove that these IDs stabilize the linear cluster state, we select the specific separable $N$-qubit IDs shown in Supplementary Figure 6 from the stabilizer group of $(H|0\rangle)^{\otimes N}$. Acting a controlled-$Z$ gate on every nearest-neighbor pair of qubits in the array transforms this separable ID into the corresponding benchmark ID from Fig. 2 of the main text, Q.E.D.

---

[*] Corresponding Author: dressel@chapman.edu; Keck Center for Science and Engineering, Chapman University, One University Drive, Orange, CA 92866; Phone: (714) 516-5949

**SUPPLEMENTARY FIGURES**

$N = 3$, $M = 4$

| $\lambda_i$ | $O_i$ |
|---|---|
| -1 | Y X Y |
| 1 | Y Y Z |
| 1 | Z X Z |
| 1 | Z Y Y |

$N = 4$, $M = 5$

| $\lambda_i$ | $O_i$ |
|---|---|
| -1 | Y X Y Z |
| 1 | Y Y Z I |
| 1 | Z X I X |
| 1 | Z Y Y Z |
| 1 | I I Z X |

$N = 5$, $M = 5$

| $\lambda_i$ | $O_i$ |
|---|---|
| -1 | Z Y X Y Z |
| 1 | Z Y Y Z I |
| 1 | I Z X I X |
| 1 | X I Y Y Z |
| 1 | X Z I Z X |

$N = 6$, $M = 5$

| $\lambda_i$ | $O_i$ |
|---|---|
| -1 | I Z Y X Y Z |
| 1 | X I Y Y Z I |
| 1 | X Z Z X I X |
| 1 | Z X I Y Y Z |
| 1 | Z X Z I Z X |

$N = 33$, $M = 10$

| $\lambda_i$ | $O_i$ |
|---|---|
| -1 | I I I Z Y Y I X Z I I I I I I Z Y X Y Z Z Y Y Z I I I I I I I I I |
| 1 | I I I I I I I I Z X I Y Y I X I Y Y Z I I I I I I I Z X Z I I I |
| 1 | I I I I I Z Y Y I X Z I I Z X Z Z X I X Z I I I I I I I I I I I |
| -1 | Z X Z I I I I I I Z Y X Y Z Z X I Y Y I X Z I I I Z X Z I I I I I |
| 1 | Z Y Y I X Z I I I I I I Z X I X Z I Z X I Y X X Y Z I I I I I I I |
| 1 | I I I I I I I I I Z X I X Z I I I I Z X I Y X X Y Z I I Z Y Y Z |
| 1 | I I I I I I Z Y Y I Y Y Z I I I I I I I Z X I X I X I Y Y I X Z |
| 1 | I I Z Y Y I Y X Y Z I I I I I I I I I I I I I Z Y Y I Y Y I Y X Y |
| 1 | Y Y I Y X Y Z I I I I I I I I I I I I I I I I I I Z Y X Y Z I I |
| -1 | Y X Y Z I I I I I I I I I I I I I I I I I I I I I I I I I Z Y Y |

$N = 32$, $M = 10$

| $\lambda_i$ | $O_i$ |
|---|---|
| -1 | I I Z Y Y I X Z I I I I I I Z Y X Y Z Z Y Y Z I I I I I I I I I |
| 1 | I I I I I I I Z X I Y Y I X I Y Y Z I I I I I I I Z X Z I I I |
| 1 | I I I I Z Y Y I X Z I I Z X Z Z X I X Z I I I I I I I I I I I |
| -1 | X Z I I I I I I Z Y X Y Z Z X I Y Y I X Z I I I Z X Z I I I I I |
| 1 | Y Y I X Z I I I I I I Z X I X Z I Z X I Y X X Y Z I I I I I I I |
| 1 | I I I I I I I I Z X I X Z I I I I Z X I Y X X Y Z I I Z X Z I |
| 1 | I I I I I Z Y Y I Y Y Z I I I I I I I Z X I X I X I Y Y Z Z X |
| -1 | I Z Y Y I Y X Y Z I I I I I I I I I I I I I Z Y Y I Y Y I X I X |
| 1 | X I Y X Y Z I I I I I I I I I I I I I I I I I I Z Y X Y I X Z |
| 1 | Y Y Z I I I I I I I I I I I I I I I I I I I I I I I I I Z X Z |

$N = 31$, $M = 10$

| $\lambda_i$ | $O_i$ |
|---|---|
| -1 | I Z Y Y I X Z I I I I I I Z Y X Y Z Z Y Y Z I I I I I I I I I |
| 1 | I I I I I I Z X I Y Y I X I Y Y Z I I I I I I I Z X Z I I I |
| 1 | I I I Z Y Y I X Z I I Z X Z Z X I X Z I I I I I I I I I I I |
| -1 | I I I I I I Z Y X Y Z Z X I Y Y I X Z I I I Z X Z I I I I I |
| 1 | X I X Z I I I I I I Z X I X Z I Z X I Y X X Y Z I I I I I I I |
| 1 | I I I I I I I Z X I X Z I I I I Z X I Y X X Y Z I I Z X Z I |
| 1 | I I I I Z Y Y I Y Y Z I I I I I I I Z X I X I X I Y Y Z Z X |
| -1 | Z Y Y I Y X Y Z I I I I I I I I I I I I I Z Y Y I Y Y I X I X |
| 1 | Z Y X Y Z I I I I I I I I I I I I I I I I I I Z Y X Y I X Z |
| 1 | X Z I I I I I I I I I I I I I I I I I I I I I I I I I Z X Z |

Supplementary Figure 1. Known minimal benchmark IDs for qubit numbers $3 \leq N \leq 33$. We present all IDs in Supplementary Figures 1–5, in ascending order of $N$ in the left column, and descending in the right column. Left: $N = 3, 4, 5, 6$. Right: $N = 31, 32, 33$.

| $\lambda_i$ | $O_i$ |
|---|---|
| | $N = 7, M = 6$ |
| 1 | I I Z Y X X Y |
| 1 | Z X I Y Y Z I |
| 1 | Z X Z Z X I X |
| -1 | I Z X I Y X Y |
| 1 | X I X Z I Z X |
| 1 | X Z I I I I I |

| $\lambda_i$ | $O_i$ |
|---|---|
| | $N = 8, M = 6$ |
| 1 | I I I Z Y X X Y |
| 1 | I Z X I Y Y Z I |
| 1 | X I X Z Z X I X |
| -1 | X Z Z X I Y X Y |
| 1 | Z X I X Z I Z X |
| 1 | Z X Z I I I I I |

| $\lambda_i$ | $O_i$ |
|---|---|
| | $N = 9, M = 6$ |
| -1 | I I I Z Y X Y Z I |
| 1 | X I X I Y Y Z I I |
| 1 | I Z X Z Z X I X Z |
| 1 | X Z Z X I Y Y I X |
| 1 | Z X I X Z I Z X Z |
| 1 | Z X Z I I I I Z X |

| $\lambda_i$ | $O_i$ |
|---|---|
| | $N = 10, M = 6$ |
| -1 | I I I Z Y X Y Z Z X |
| 1 | X I X I Y Y Z I I I |
| 1 | I Z X Z Z X I X Z I |
| 1 | X Z Z X I Y Y I X Z |
| 1 | Z X I X Z I Z X I X |
| 1 | Z X Z I I I I Z X Z |

| $\lambda_i$ | $O_i$ |
|---|---|
| | $N = 30, M = 10$ |
| -1 | Z Y Y I X Z I I I I I I Z Y X Y Z Z Y Y Z I I I I I I I I I |
| 1 | I I I I Z X I Y Y I X I Y Y Z I I I I I I I I Z X Z I I I |
| 1 | I I Z Y Y I X Z I I Z X Z Z X I X Z I I I I I I I I I I I |
| -1 | I I I I I I Z Y X Y Z Z X I Y Y I X Z I I I Z X Z I I I I I |
| 1 | Z X Z I I I I I I Z X I X Z I Z X I Y X X Y Z I I I I I I I |
| 1 | I I I I I I I Z X I X Z I I I I Z X I Y X X Y Z I I Z X Z I |
| 1 | I I I Z Y Y I Y Y Z I I I I I I I I Z X I X I X I Y Y Z Z X |
| -1 | Y Y I Y X Y Z I I I I I I I I I I I I I Z Y Y I Y Y I X I X |
| 1 | Y X Y Z I I I I I I I I I I I I I I I I I I I I Z Y X Y I X Z |
| 1 | I I I I I I I I I I I I I I I I I I I I I I I I I I I Z X Z |

| $\lambda_i$ | $O_i$ |
|---|---|
| | $N = 29, M = 10$ |
| -1 | Z Y Y I X Z I I I I I I Z Y X Y Z Z Y Y Z I I I I I I I I |
| 1 | I I I I Z X I Y Y I X I Y Y Z I I I I I I I I Z X Z I I |
| 1 | I I Z Y Y I X Z I I Z X Z Z X I X Z I I I I I I I I I I I |
| -1 | I I I I I I Z Y X Y Z Z X I Y Y I X Z I I I Z X Z I I I I |
| 1 | Z X Z I I I I I I Z X I X Z I Z X I Y X X Y Z I I I I I I |
| 1 | I I I I I I I Z X I X Z I I I I Z X I Y X X Y Z I I Z X Z |
| 1 | I I I Z Y Y I Y Y Z I I I I I I I I Z X I X I X I Y Y Z I |
| -1 | Y Y I Y X Y Z I I I I I I I I I I I I I Z Y Y I Y Y I X Z |
| 1 | Y X Y Z I I I I I I I I I I I I I I I I I I I I Z Y X Y I X |
| 1 | I I I I I I I I I I I I I I I I I I I I I I I I I I I Z X |

| $\lambda_i$ | $O_i$ |
|---|---|
| | $N = 28, M = 9$ |
| -1 | Z Y Y I X Z I I I I I I Z Y X Y Z Z Y Y Z I I I I I I I |
| 1 | I I I I Z X I Y Y I X I Y Y Z I I I I I I I I Z X Z I |
| 1 | I I Z Y Y I X Z I I Z X Z Z X I X Z I I I I I I I I I I |
| -1 | I I I I I I Z Y X Y Z Z X I Y Y I X Z I I I Z X Z I I I |
| 1 | Z X Z I I I I I I Z X I X Z I Z X I Y X X Y Z I I I I I |
| 1 | I I I I I I I Z X I X Z I I I I Z X I Y X X Y Z I I Z X |
| 1 | I I I Z Y Y I Y Y Z I I I I I I I I Z X I X I X I Y Y Z |
| -1 | Y Y I Y X Y Z I I I I I I I I I I I I I Z Y Y I Y Y I X |
| 1 | Y X Y Z I I I I I I I I I I I I I I I I I I I I Z Y X Y Z |

Supplementary Figure 2. Known minimal benchmark IDs for qubit numbers $3 \leq N \leq 33$. We present all IDs in Supplementary Figures 1–5, in ascending order of $N$ in the left column, and descending in the right column. Left: $N = 7, 8, 9, 10$. Right: $N = 28, 29, 30$.

$N = 11$, $M = 7$

| $\lambda_i$ | $O_i$ |
|---|---|
| -1 | I I I I Z Y X Y Z Z X Z |
| 1 | X I X I Y Y Z I I I I |
| 1 | I Z X Z Z X I X Z I I |
| 1 | X Z Z X I Y Y I X Z I |
| 1 | Z X I X Z I Z X I X Z |
| 1 | Z X Z I I I I Z X I X |
| 1 | I I I I I I I I I Z X |

$N = 12$, $M = 7$

| $\lambda_i$ | $O_i$ |
|---|---|
| -1 | I I I I Z Y X Y Z Z X Z |
| 1 | Y Y I X I Y Y Z I I I I |
| 1 | I I Z X Z Z X I X Z I I |
| 1 | Y Y Z Z X I Y Y I X Z I |
| 1 | I Z X I X Z I Z X I X Z |
| 1 | X I X Z I I I I Z X I X |
| 1 | X Z I I I I I I I I Z X |

$N = 13$, $M = 7$

| $\lambda_i$ | $O_i$ |
|---|---|
| -1 | I I I I Z Y X Y Z Z X Z I |
| 1 | Y Y I X I Y Y Z I I I I I |
| 1 | I I Z X Z Z X I X Z I I I |
| 1 | Y Y Z Z X I Y Y I X Z Z X |
| 1 | I Z X I X Z I Z X I X I X |
| 1 | X I X Z I I I I Z X I X Z |
| 1 | X Z I I I I I I I I Z X Z |

$N = 14$, $M = 7$

| $\lambda_i$ | $O_i$ |
|---|---|
| -1 | I I I I I Z Y X Y Z Z Y Y Z |
| 1 | Z Y Y I X I Y Y Z I I I I I |
| 1 | I I I Z X Z Z X I X Z I I I |
| -1 | Y X Y Z Z X I Y Y I X Z I I |
| -1 | I I Z X I X Z I Z X I Y X Y |
| 1 | Z X I X Z I I I I Z X I X Z |
| 1 | Y Y Z I I I I I I I I Z Y Y |

$N = 27$, $M = 9$

| $\lambda_i$ | $O_i$ |
|---|---|
| -1 | Z X I X Z I I I I I I Z Y X Y Z Z Y Y Z I I I I I I I |
| 1 | I I I I Z X I Y Y I X I Y Y Z I I I I I I I I I Z X Z I |
| -1 | Z Y X Y I X Z I I Z X Z Z X I X Z I I I I I I I I I I |
| -1 | I I I I Z Y X Y Z Z X I Y Y I X Z I I I Z X Z I I I |
| 1 | I I I I I I I I Z X I X Z I Z X I Y X X Y Z I I I I I |
| 1 | I I I I I I Z X I X Z I I I I Z X I Y X X Y Z I I Z X |
| 1 | I I Z Y Y I Y Y Z I I I I I I I I I Z X I X I X I Y Y Z |
| -1 | Y X X X Y Z I I I I I I I I I I I I I I Z Y Y I Y Y I X |
| -1 | Y Y Z I I I I I I I I I I I I I I I I I I I I I Z Y X Y Z |

$N = 26$, $M = 9$

| $\lambda_i$ | $O_i$ |
|---|---|
| -1 | Z X I X Z I I I I I I Z Y X Y Z Z Y Y Z I I I I I I |
| 1 | I I I I Z X I Y Y I X I Y Y Z I I I I I I I I I Z X Z |
| -1 | Z Y X Y I X Z I I Z X Z Z X I X Z I I I I I I I I I |
| -1 | I I I I Z Y X Y Z Z X I Y Y I X Z I I I Z X Z I I |
| 1 | I I I I I I I I Z X I X Z I Z X I Y X X Y Z I I I I |
| 1 | I I I I I I Z X I X Z I I I I Z X I Y X X Y Z I I I |
| 1 | I I Z Y Y I Y Y Z I I I I I I I I I Z X I X I X I Y Y |
| -1 | Y X X X Y Z I I I I I I I I I I I I I I Z Y Y I Y Y Z |
| -1 | Y Y Z I I I I I I I I I I I I I I I I I I I I I Z Y X Y |

$N = 25$, $M = 9$

| $\lambda_i$ | $O_i$ |
|---|---|
| -1 | I I Z X Z I I I I I I I Z Y X Y Z Z Y Y Z I I I I I |
| 1 | I I I I Z X I Y Y I X I Y Y Z I I I I I I I I I I |
| 1 | X I Y Y I X Z I I Z X Z Z X I X Z I I I I I I I I |
| -1 | I I I I I Z Y X Y Z Z X I Y Y I X Z I I I Z X Z I |
| 1 | X Z I I I I I I Z X I X Z I Z X I Y X X Y Z I I I |
| 1 | I I I I I I Z X I X Z I I I I Z X I Y X X Y Z Z X |
| 1 | Z X I Y Y I Y Y Z I I I I I I I I I Z X I X I X I X |
| -1 | I Z Y X Y Z I I I I I I I I I I I I I I I Z Y Y I X Z |
| 1 | Z X Z I I I I I I I I I I I I I I I I I I I I Z X Z |

Supplementary Figure 3. Known minimal benchmark IDs for qubit numbers $3 \leq N \leq 33$. We present all IDs in Supplementary Figures 1–5, in ascending order of $N$ in the left column, and descending in the right column. Left: $N = 11, 12, 13, 14$. Right: $N = 25, 26, 27$.

### $N = 15, M = 7$

| $\lambda_i$ | $O_i$ |
|---|---|
| -1 | I I I I I Z Y X Y Z Z Y Y Z I |
| 1 | Z Y Y I X I Y Y Z I I I I I I |
| 1 | I I I Z X Z Z X I X Z I I Z X |
| -1 | Y X Y Z Z X I Y Y I X Z I I I |
| -1 | I I Z X I X Z I Z X I Y X Y Z |
| 1 | Z X I X Z I I I I Z X I X I X |
| 1 | Y Y Z I I I I I I I I Z Y Y Z |

### $N = 16, M = 8$

| $\lambda_i$ | $O_i$ |
|---|---|
| -1 | I I I I I Z Y X Y Z Z Y Y Z I I |
| 1 | Z Y Y I X I Y Y Z I I I I I I I |
| 1 | I I I Z X Z Z X I X Z I I Z X Z |
| -1 | Y X Y Z Z X I Y Y I X Z I I I I |
| -1 | I I Z X I X Z I Z X I Y X Y Z I |
| 1 | Z X I X Z I I I I Z X I X I X Z |
| 1 | Y Y Z I I I I I I I I Z Y Y I X |
| 1 | I I I I I I I I I I I I I I Z X |

### $N = 17, M = 8$

| $\lambda_i$ | $O_i$ |
|---|---|
| -1 | I I I I I Z Y X Y Z Z Y Y Z I I I |
| 1 | Z Y Y I X I Y Y Z I I I I I I I I |
| 1 | I I I Z X Z Z X I X Z I I Z X Z I |
| -1 | Y X Y Z Z X I Y Y I X Z I I I I I |
| -1 | I I Z X I X Z I Z X I Y X Y Z Z X |
| 1 | Z X I X Z I I I I Z X I X I X I X |
| 1 | Y Y Z I I I I I I I I Z Y Y I X Z |
| 1 | I I I I I I I I I I I I I I Z X Z |

### $N = 24, M = 9$

| $\lambda_i$ | $O_i$ |
|---|---|
| -1 | I Z X Z I I I I I I Z Y X Y Z Z Y Y Z I I I I I |
| 1 | I I I Z X I Y Y I X I Y Y Z I I I I I I I I I I |
| 1 | Z Y Y I X Z I I Z X Z Z X I X Z I I I I I I I I |
| -1 | I I I I Z Y X Y Z Z X I Y Y I X Z I I I Z X Z I |
| 1 | I I I I I I I Z X I X Z I Z X I Y X X Y Z I I I |
| 1 | I I I I Z X I X Z I I I I Z X I Y X X Y Z Z X |
| 1 | X I Y Y I Y Y Z I I I I I I I I Z X I X I X I X |
| -1 | Z Y X Y Z I I I I I I I I I I I I I I Z Y Y I X Z |
| 1 | X Z I I I I I I I I I I I I I I I I I I I I Z X Z |

### $N = 23, M = 9$

| $\lambda_i$ | $O_i$ |
|---|---|
| -1 | Z X Z I I I I I I Z Y X Y Z Z Y Y Z I I I I I |
| 1 | I I Z X I Y Y I X I Y Y Z I I I I I I I I I I |
| 1 | Y Y I X Z I I Z X Z Z X I X Z I I I I I I I I |
| -1 | I I I Z Y X Y Z Z X I Y Y I X Z I I I Z X Z I |
| 1 | I I I I I I Z X I X Z I Z X I Y X X Y Z I I I |
| 1 | I I I I Z X I X Z I I I I Z X I Y X X Y Z Z X |
| 1 | Z Y Y I Y Y Z I I I I I I I I I Z X I X I X I X |
| -1 | Y X Y Z I I I I I I I I I I I I I I I Z Y Y I X Z |
| 1 | I I I I I I I I I I I I I I I I I I I I I Z X Z |

### $N = 22, M = 9$

| $\lambda_i$ | $O_i$ |
|---|---|
| -1 | Z X Z I I I I I I Z Y X Y Z Z Y Y Z I I I I |
| 1 | I I Z X I Y Y I X I Y Y Z I I I I I I I I I |
| 1 | Y Y I X Z I I Z X Z Z X I X Z I I I I I I I |
| -1 | I I I Z Y X Y Z Z X I Y Y I X Z I I I Z X Z |
| 1 | I I I I I I Z X I X Z I Z X I Y X X Y Z I I |
| 1 | I I I I Z X I X Z I I I I Z X I Y X X Y Z I |
| 1 | Z Y Y I Y Y Z I I I I I I I I I Z X I X I X Z |
| -1 | Y X Y Z I I I I I I I I I I I I I I I Z Y Y I X |
| 1 | I I I I I I I I I I I I I I I I I I I I I Z X |

Supplementary Figure 4. Known minimal benchmark IDs for qubit numbers $3 \leq N \leq 33$. We present all IDs in Supplementary Figures 1–5, in ascending order of $N$ in the left column, and descending in the right column. Left: $N = 15, 16, 17$. Right: $N = 22, 23, 24$.

### $N = 18, M = 8$

| $\lambda_i$ | $O_i$ |
|---|---|
| -1 | I I I I I I Z Y X Y Z Z Y Y Z I I I |
| 1 | I Z Y Y I X I Y Y Z I I I I I I I I |
| 1 | I I I I Z X Z Z X I X Z I I Z X Z I |
| -1 | Z Y X Y Z Z X I Y Y I X Z I I I I I |
| -1 | I I I Z X I X Z I Z X I Y X Y Z Z X |
| 1 | X I X I X Z I I I I Z X I X I X I X |
| 1 | Z Y Y Z I I I I I I I I Z Y Y I X Z |
| 1 | X Z I I I I I I I I I I I I I Z X Z |

### $N = 21, M = 8$

| $\lambda_i$ | $O_i$ |
|---|---|
| -1 | Z X Z I I I I I I I Z Y X Y Z Z Y Y Z I I I |
| 1 | I I Z X I Y Y I X I Y Y Z I I I I I I I I |
| 1 | Y Y I X Z I I Z X Z Z X I X Z I I I I I I |
| -1 | I I I Z Y X Y Z Z X I Y Y I X Z I I I Z X |
| 1 | I I I I I I Z X I X Z I Z X I Y X X Y Z I |
| 1 | I I I I Z X I X Z I I I I Z X I Y X X Y Z |
| 1 | Z Y Y I Y Y Z I I I I I I I I Z X I X I X |
| -1 | Y X Y Z I I I I I I I I I I I I I I Z Y Y Z |

### $N = 19, M = 8$

| $\lambda_i$ | $O_i$ |
|---|---|
| -1 | I I I I I I I Z Y X Y Z Z Y Y Z I I I |
| 1 | I I Z Y Y I X I Y Y Z I I I I I I I I |
| 1 | I I I I I Z X Z Z X I X Z I I Z X Z I |
| -1 | I Z Y X Y Z Z X I Y Y I X Z I I I I I |
| -1 | X Z I I Z X I X Z I Z X I Y X Y Z Z X |
| 1 | Z X I X I X Z I I I I Z X I X I X I X |
| 1 | X I Y Y Z I I I I I I I I Z Y Y I X Z |
| 1 | Z X Z I I I I I I I I I I I I I I Z X Z |

### $N = 20, M = 8$

| $\lambda_i$ | $O_i$ |
|---|---|
| -1 | I I I I I I I I I Z Y X Y Z Z Y Y I X Z |
| 1 | Z X I X I Y Y I X I Y Y Z I I I I I I I |
| 1 | I I Z X Z I I Z X Z Z X I X Z I I I I I |
| -1 | I I I Z Y X Y Z Z X I Y Y I X Z I I I I |
| 1 | I I I I I I Z X I X Z I Z X I Y X X Y Z |
| 1 | Y Y Z I Z X I X Z I I I I Z X I Y X X Y |
| 1 | Z Y Y I Y Y Z I I I I I I I I Z X Z I I |
| -1 | Y X Y Z I I I I I I I I I I I I I I Z Y Y |

Supplementary Figure 5. Known minimal benchmark IDs for qubit numbers $3 \leq N \leq 33$. We present all IDs in Supplementary Figures 1–5, in ascending order of $N$ in the left column, and descending in the right column. Left: $N = 18, 19$. Right: $N = 20, 21$.

| $\lambda_i$ | $O_i$ |
|---|---|
| 1 | X X X X I X ⋯ ⋯ I X I X |
| 1 | I I X I X I ⋯ ⋯ X I X X |
| 1 | X I X X I I ⋯ ⋯ I I I I |
| -1 | X X X I I I ⋯ ⋯ I I I I |
| 1 | X I I I I I ⋯ ⋯ I I I I |
| 1 | I I I I X I ⋱    I I I I |
| 1 | I I I I I X I ⋱    I I I |
| ⋮ | ⋮ ⋮ ⋮ ⋮ ⋱ ⋱ ⋱ ⋱ ⋮ ⋮ |
| ⋮ | ⋮ ⋮ ⋮ ⋮   ⋱ ⋱ ⋱ ⋱ ⋮ ⋮ |
| 1 | I I I I I I ⋱ I X I I I |
| 1 | I I I I I I I ⋱ I X I I |
| 1 | I I I I I I ⋯ ⋯ I I X X |
| 1 | I I I I I I ⋯ ⋯ I I I X |

(a)

| $\lambda_i$ | $O_i$ |
|---|---|
| 1 | X X X X I X ⋯ ⋯ I X I X X |
| 1 | I I X I X I ⋯ ⋯ X I X I X |
| 1 | X I X X I I ⋯ ⋯ I I I I I |
| -1 | X X X I I I ⋯ ⋯ I I I I I |
| 1 | X I I I I I ⋯ ⋯ I I I I I |
| 1 | I I I I X I ⋱   I I I I I |
| 1 | I I I I I X I ⋱   I I I I |
| ⋮ | ⋮ ⋮ ⋮ ⋮ ⋱ ⋱ ⋱ ⋱ ⋮ ⋮ ⋮ |
| ⋮ | ⋮ ⋮ ⋮ ⋮   ⋱ ⋱ ⋱ ⋱ ⋮ ⋮ ⋮ |
| 1 | I I I I I I ⋱ I X I I I I |
| 1 | I I I I I I I ⋱ I X I I I |
| 1 | I I I I I I ⋯ ⋯ I I X I I |
| 1 | I I I I I I ⋯ ⋯ I I I X X |
| 1 | I I I I I I ⋯ ⋯ I I I I X |

(b)

Supplementary Figure 6. Separable $N$-qubit IDs from the stabilizer group of $(H|0\rangle)^{\otimes N}$, for (a) even $N \geq 10$ and (b) odd $N \geq 11$. These IDs can be extended in increments of two qubits and two observables by adding tiles as shown, and filling all other spaces with 'I's. Acting a controlled-$Z$ gate on every nearest-neighbor pair of qubits in the array transforms these separable IDs into the corresponding benchmark ID from Fig. 2 of the main text.



\end{document}